\documentclass{ws-procs10x7}

\begin{document}

\title{Searches for Supersymmetry at the Tevatron}

\author{Arnd Meyer}

\address{III.~Phys.~Inst.~A, RWTH Aachen, 52074 Aachen, Germany \\
E-mail: meyera@fnal.gov \\
(for the D\O\ and CDF Collaborations)}

\def \ud {{1 \over 2} }
\newcommand{\chiO}{$\tilde{\chi}_0^1$}
\newcommand{\rpv}{\mbox{$\not \hspace{-0.10cm} R_p$}}
\newcommand{\MET}{\mbox{$\not \hspace{-0.10cm} E_T$}}

\twocolumn[\maketitle\abstract{
Both Tevatron experiments, D\O\ and CDF, have searched for signs of Supersymmetry in the
present Run II data sample, using integrated luminosities of up to 260~pb$^{-1}$ collected
in $p\bar{p}$ collisions at a center-of-mass energy of $1.96$~TeV. In these proceedings, new
results are presented in the search for squarks and gluinos in the jets and missing
transverse energy final state, associated production of charginos and neutralinos with
multilepton final states, search for the rare decay $B_s \rightarrow \mu\mu$, searches
allowing R-parity violation (muons+jets, multileptons), and searches in the gauge mediated
supersymmetry breaking framework with the final state of two photons and missing transverse
energy. In the absence of any significant deviation from Standard Model expectations, limits
on the presence of new physics are set, which in many cases are the most stringent to date.
}]

After the very successful Tevatron Run I, the accelerator and the
colliding beam experiments CDF and D\O\ were upgraded for Run II. On the accelerator side,
the beam energy has been increased from $900\:\mbox{GeV}$ to $980\:\mbox{GeV}$. A new
proton storage ring, the ``Main Injector'', has been built, and the same tunnel houses
the ``Recycler'' storage ring, which will improve the rate at which antiprotons can be
accumulated. The bunch spacing in the Tevatron has been reduced from $3.5\:\mu\mbox{s}$
to $396\:\mbox{ns}$. The results shown below\cite{prelim} are based on early Run II
data collected in 2002-2004 and correspond to an integrated luminosity of up to
$260\:\mbox{pb}^{-1}$, significantly larger than the Run I data set
($\int {\cal L}\, dt \simeq 110\:\mbox{pb}^{-1}$).

\section{Search for New Physics with Jets \boldmath  $+$ \MET \unboldmath }

\begin{figure}
  \epsfxsize6.0cm
  \figurebox{}{}{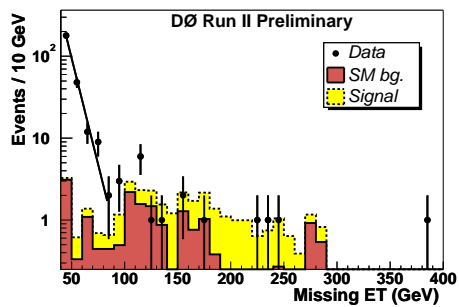}
  \caption{Distribution of \MET\ in the D\O\ acoplanar jets $+$ \MET\ analysis
           to search for squarks and gluinos.}
  \label{squarks1}
  \end{figure}

  Both CDF and D\O\ have reported preliminary results in signatures with jets and \MET.
  In the light quark jets $+$ \MET\ search based on $85\:\mbox{pb}^{-1}$ of D\O\ data,
  two high $E_T$ jets and large $H_T > 275\:\mbox{GeV}$ are required, and \MET\
  $>175\:\mbox{GeV}$ (Fig.~\ref{squarks1}). It is worth noting the single event in the
  tail of the \MET\ distribution: it has four jets with $E_T$ of $289\:\mbox{GeV}$,
  $117\:\mbox{GeV}$, $14\:\mbox{GeV}$, and $11\:\mbox{GeV}$, and \MET\ $=
  381\:\mbox{GeV}$. After all cuts, 4 events remain in the data, with $2.67 \pm 0.95$
  expected from Standard Model processes. The dominant background is due to $Z
  \rightarrow \nu\nu + 2$ jets. Limits are set on the production of light squarks and
  gluinos in an mSUGRA scenario ($m_0 = 25\:\mbox{GeV}$, $A_0 = 0$, $\tan\beta = 3$, $\mu
  < 0$): $292\:\mbox{GeV}$ for squarks, and $333\:\mbox{GeV}$ for gluinos, beyond the Run
  I excluded domain.

\begin{figure}
  \epsfxsize4.6cm
  \figurebox{80pt}{80pt}{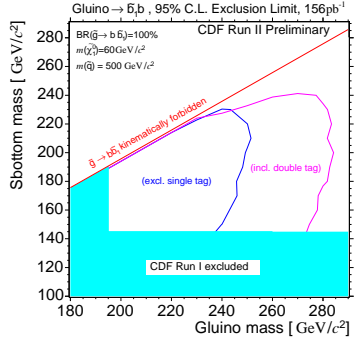}
  \caption{The 95\% C.L. exclusion region from the CDF four jet $+$ \MET\ search
           for gluino pair production and decay into sbottom squarks.}
  \label{sbottom2}
  \end{figure}

  CDF has searched for pair production of gluinos and decays gluino $\rightarrow$
  sbottom $+$ bottom in $156\:\mbox{pb}^{-1}$ of data. The analysis requires four
  jets and large \MET, and one or two b-tags. In the more sensitive double tag analysis,
  4 events are observed with \MET\ $> 80\:\mbox{GeV}$, where
  $2.6\pm 0.7$ events are expected from Standard Model processes (mostly top quark
  pair production). The exclusion limits, significantly extending earlier results,
  in the gluino sbottom mass plane are shown in Fig.~\ref{sbottom2}.

\section{Search for \boldmath $B_s^0 \rightarrow \mu^+\mu^-$ \unboldmath }

  The purely leptonic decays $B_{d,s}^0 \rightarrow \mu^+
  \mu^-$ are flavor-changing neutral current processes.
  In the Standard Model, these decays are forbidden at the tree level and
  proceed at a very low rate through higher-order diagrams.
  However, the decay amplitude can be
  enhanced in some extensions of the Standard Model. In the MSSM for example
  ${\cal B}(B^0_s \rightarrow \mu^+ \mu^-) \propto(\tan\beta)^6$, leading to an
  enhancement of up to three orders of magnitude.

  Both D\O\ and CDF have reported new results
  of a search for the decay $B^0_s \rightarrow \mu^+ \mu^-$.
  Using a data set with integrated luminosity of 240~pb$^{-1}$, D\O\ sets an upper
  limit on the branching fraction of
  ${\cal B}(B^0_s \rightarrow \mu^+ \mu^-) \leq 5.0\times 10^{-7}$ at the 95\% C.L.,
  the most stringent upper bound to date.
  Based on 171~pb$^{-1}$ of data, CDF determined
  ${\cal B}(B^0_s \rightarrow \mu^+\mu^-)<7.5\times 10^{-7}$ at the 
  95\% C.L.\cite{bscdfII}.

\section{Associated Production of Charginos and Neutralinos}

\begin{figure}
  \epsfxsize5.0cm
  \figurebox{}{}{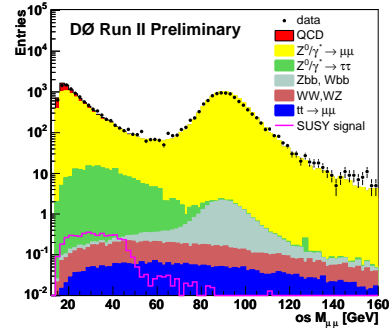}
  \caption{Invariant dimuon mass spectrum in the D\O\ search for associated production of charginos
           and neutralinos.}
  \label{trilep_m}
  \end{figure}

  The three-lepton signature ($+$ \MET) is a gold plated signature for
  the associated production of charginos and neutralinos. D\O\ has
  presented four different analyses, in the end combined for
  maximum sensitivity. The four analyses are based on the signature of
  two electrons plus an isolated track, two muons plus an isolated
  track, one electron plus one muon plus an isolated track, and
  finally two muons with the same charge. The isolated track requirement
  is sensitive to electrons, muons, and taus, and maximizes efficiency
  by not requiring explicit lepton identification. All identified
  leptons are required to be isolated. The invariant dimuon mass spectrum
  is shown in Fig.~\ref{trilep_m}, illustrating background and expected
  signal contributions. Challenges for the analyses are the
  very small signal cross sections ($\sigma \times \mbox{BR} < 0.5\:\mbox{pb}$)
  and typically low lepton momenta due to cascade decays. The data samples
  used correspond to $147-249\:\mbox{pb}^{-1}$. After all cuts, 1, 1, 0,
  and 1 event remain in the data, compatible with the Standard Model
  expectations of $0.7\pm 0.5$, $1.8 \pm 0.5$, $0.3 \pm 0.3$, and
  $0.1 \pm 0.1$ events. 

\begin{figure}
  \epsfxsize6.0cm
  \figurebox{}{}{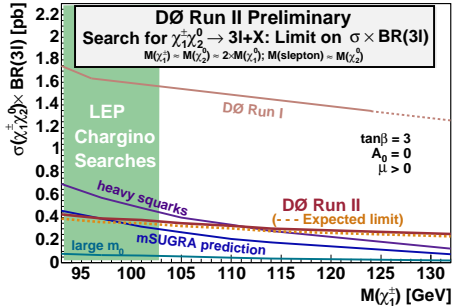}
  \caption{Combined D\O\ limits from the search for associated production of charginos
           and neutralinos.}
  \label{trilep}
  \end{figure}

  The derived cross section limits, shown in Fig.~\ref{trilep}, are significantly
  more stringent than Run I results. The corresponding chargino mass limit is close
  to the LEP limits, for an mSUGRA scenario with maximum leptonic branching fractions .

\section{R-parity Violation}

  Supersymmetry does not impose the conservation of lepton and baryon numbers. Allowing
  R-parity to be violated leads to signatures distinct from R-parity conserving
  models.
  D\O\ has performed searches for not vanishing $LL\bar{E}$ couplings $\lambda_{121}$, 
  $\lambda_{122}$ and $LQ\bar{D}$ coupling $\lambda'_{211}$. For non-zero $LL\bar{E}$
  coupling SUSY particles are produced in pairs or associated, and (cascade) decay to the
  lightest neutralino \chiO, which then decays into two charged leptons and one neutrino by
  violating R-parity. The final state includes four charged leptons (electrons or muons
  for $\lambda_{121}$ and $\lambda_{122}$) and two neutrinos which lead to missing
  transverse energy.

  The $LQ\bar{D}$ coupling offers the opportunity to produce the scalar supersymmetric
  particles as resonances. For $\lambda'_{211}$ this is either a smuon or a muon
  sneutrino. The smuon can decay into the lightest neutralino \chiO\ and a high $p_T$
  muon. The neutralino decays via the same R-parity violating coupling $\lambda'_{211}$
  into one muon and two jets.

  \subsection{R-parity Violation via $\lambda_{121}$ and $\lambda_{122}$
              ($eel$ and $\mu\mu l$ Final States)}

    The data samples correspond to an integrated luminosity of 238~pb$^{-1}$
    ($\lambda_{121}$) and 160~pb$^{-1}$ ($\lambda_{122}$). For the $\lambda_{121}$ analysis,
    the identification of two electrons and a third lepton, an electron or a muon, is
    required. A fourth lepton is not required. The Z resonance is cut away and significant
    \MET\ is required. After all cuts no data events remain while $0.45\pm 0.43$ events are
    expected from Standard Model processes.

    For the $\lambda_{122}$ analysis, events with at least two muons plus an electron or a muon
    are selected. No fourth lepton is required for maximum efficiency, similar to the electron
    channel. Two-dimensional cuts in the invariant dimuon mass, \MET, and the sum of muon $p_T$
    are applied to improve the sensitivity. Two events are found in the data, while
    $0.63\pm 1.93$ events are expected from Standard Model processes.

  \subsection{\mbox{R-parity Violation via $\lambda'_{211}$ ($\mu\mu$ jet jet)}}

    The search in the resonant channel uses data with a luminosity of 154~pb$^{-1}$. The data
    sample contains events with two well measured muons and two jets. The slepton and
    neutralino masses are reconstructed, since all decay products are detected. The neutralino
    mass is calculated using the two jets and the next to leading muon, since the muon created
    at the slepton decay vertex together with the neutralino is with high propability the
    leading $p_T$ muon. Sliding cuts on the leading muon transverse momentum and the
    reconstructed Z, \chiO\ and $\tilde{l}$ masses are applied depending on the neutralino and
    slepton mass of the SUSY parameter point under study. No excess above the Standard Model
    expectation is observed.

  \subsection{Results}

\begin{figure}
    \epsfxsize5.5cm
    \figurebox{}{}{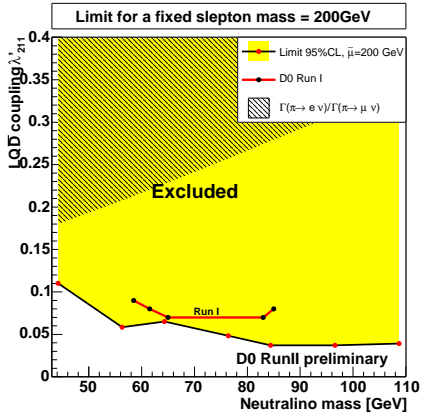}
    \caption{D\O\ search for non-vanishing $\lambda'_{211}$ coupling, 95\% C.L. limits on the lightest
             neutralino mass for a fixed $M_{\tilde{l}}=200~$GeV.}
    \label{smuon1}
    \end{figure}

    In the absence of an excess in the data, upper limits are set on the production cross
    sections. The cross section limits are translated into limits on the gaugino and slepton
    masses (see examples in Table \ref{rpvlimits} and Fig.~\ref{smuon1}).

    \begin{table}[h]
    {\begin{tabular}{rl}
    $ \lambda_{121} = 0.01 $ & $\tan \beta=5$ \\\hline
    $\mu<0$: & $M_{\tilde{\chi}^\pm}<184$~GeV \\
    $\mu>0$: & $M_{\tilde{\chi}^\pm}<181$~GeV \\\hline\hline
    $ \lambda_{122} = 0.001 $  & $\tan \beta=5$ \\\hline
    $\mu<0$: & $M_{\tilde{\chi}^\pm}<160$~GeV \\
    $\mu>0$: & $M_{\tilde{\chi}^\pm}<165$~GeV \\\hline\hline
    $ {\bf\lambda'_{211}}=0.07 $ & $\tan \beta=2$, $\mu<0$ \\\hline
    $M_{\tilde{l}}=200~$GeV:        & $~54 < M_{\tilde{\chi}^0_1} < 110$~GeV \\
    $M_{\tilde{\chi}_0^1}=~75$~GeV: & $160 < M_{\tilde{l}} < 255$~GeV  \\
    $M_{\tilde{\chi}_0^1}=100$~GeV: & $190 < M_{\tilde{l}} < 255$~GeV  \\
    \end{tabular}}
    \caption{D\O\ results on excluded gaugino and slepton masses with 95\% C.L. for the
             R-parity violating coupling parameters $\lambda_{121}$, $\lambda_{122}$ and $\lambda'_{211}$.}
    \label{rpvlimits}
    \end{table}

\section{Gauge Mediated SUSY Breaking}

\begin{figure}
  \epsfxsize4.5cm
  \figurebox{}{}{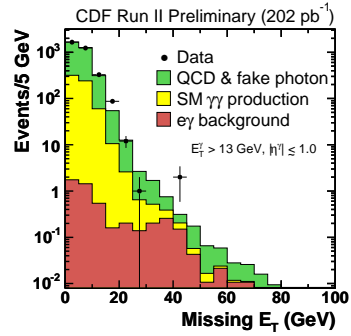}
  \caption{The \MET\ distribution for the CDF $\gamma\gamma$ data sample.}
  \label{gmsb_met}
  \end{figure}

\begin{figure}
  \epsfxsize5.2cm
  \figurebox{}{}{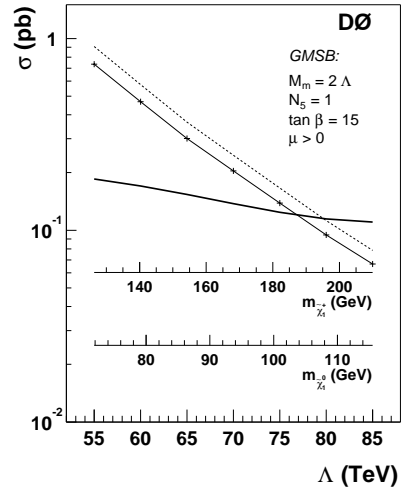}
  \caption{The 95\% C.L. exclusion region from D\O\ in the $\gamma\gamma + $ \MET\ search
           for GMSB SUSY.}
  \label{gmsb1}
  \end{figure}

  In these analyses, perfomed both by CDF and D\O\ with 202~pb$^{-1}$ and 263~pb$^{-1}$,
  respectively, the NLSP is assumed to be the lightest neutralino. The decay of the
  neutralinos into gravitino and photon leads to a final state with two photons and \MET. The
  dominant backgrounds are from direct photon production and from jets misidentified as a
  photon (Fig.~\ref{gmsb_met}). CDF (D\O ) requires two photons with $E_T > 13
  (20)\:\mbox{GeV}$ in the central part of the detector, and \MET\ $>45(40)\:\mbox{GeV}$. The
  experiments observe 0 (CDF) and 2 (D\O) events in the data, with 0.6 (CDF) and 3.7 (D\O)
  events expected in the Standard Model. For the number of messengers $N=1$, the messenger
  mass $M_m = 2\Lambda$, $\mu>0$, and $\tan\beta = 15$, the limits on the neutralino mass
  (Fig.~\ref{gmsb1}) are $93\:\mbox{GeV}$ (CDF) and $108\:\mbox{GeV}$ (D\O), extending beyond
  the LEPII limit.

\end{document}